\documentclass[aps,prd,twocolumn,amssymb,showpacs,nofootinbib]{revtex4}
\usepackage{epsf}
\usepackage{epsfig}
\usepackage{graphics}
\usepackage{graphicx}
\usepackage{amssymb}
\usepackage{amsmath}

\newcommand{\nc}{\newcommand}
\nc{\postscript}[2] 
{\setlength{\epsfxsize}{#2\hsize}\centerline{\epsfbox{#1}}}
\nc{\non}{\nonumber}
\nc{\hc}{\hbox {h.c.}} \nc{\re}{\hbox {Re}} 
\nc{\mev}{\hbox {MeV}} \nc{\gev}{\;\hbox {GeV}} \nc{\tev}{\;\hbox {TeV}}
\def\lsim{\mathrel{\raise.3ex\hbox{$<$\kern-.75em\lower1ex\hbox{$\sim$}}}}
\def\gsim{\mathrel{\raise.3ex\hbox{$>$\kern-.75em\lower1ex\hbox{$\sim$}}}}

\nc{\etal}{{\it et al.}}
\nc{\Lsp}{\;\;\;\;\;\;\;\;\;\;}  \nc{\LLLsp}{\lspace \lspace}
\nc{\lsp}{\;\;\;\;\;\;}
\nc{\spac}{\;\;\;}
\nc{\noi}{\noindent}
\nc{\beq}{\begin{equation}}   \nc{\eeq}{\end{equation}}
\nc{\bea}{\begin{eqnarray}}   \nc{\eea}{\end{eqnarray}}
\nc{\baa}{\begin{array}}      \nc{\eaa}{\end{array}}
\nc{\bit}{\begin{itemize}}    \nc{\eit}{\end{itemize}}
\nc{\ben}{\begin{enumerate}}  \nc{\een}{\end{enumerate}}
\nc{\bce}{\begin{center}}     \nc{\ece}{\end{center}}

\def\sq2{\sqrt{2}}

\def\ph{\varphi}

\def\m4{m^4(\ph)}
\def\mn2{m_n^2}

\def\v5{V^{(5)}}

\begin{document}

\title{\begin{flushright}
       \mbox{\normalsize \rm UMD-PP-08-006}\\
       \mbox{\normalsize \rm NSF-KITP-08-35}
       \end{flushright}
       \vskip 20pt\bf Odd Tachyons in Compact Extra Dimensions}
\author{Manuel Toharia\\
{\it Department of Physics, University of Maryland\\
College Park, MD 20742, USA}}

\date{\today}

\begin{abstract}
We consider a real scalar field with an arbitrary negative bulk mass term 
in a general 5D setup, where the extra spatial
coordinate is a warped interval of size $\pi R$. 
When the 5D field verifies Neumann conditions at the boundaries of the interval, the setup
will always contain at least one tachyonic KK mode. 
On the other hand, when the 5D scalar verifies Dirichlet conditions, 
there is always a critical (negative) mass $M_{c}^2$ such that the
Dirichlet scalar is stable as long as its (negative) bulk mass $\mu^2$
verifies $M^2_{c}<\mu^2$. Also, if we fix the bulk mass $\mu^2$ to a sufficiently negative
value, there will always be a critical interval distance $\pi R_c$ such that the
setup is unstable for $R>R_c$.
We point out that the best mass (or distance) bound is obtained for the
Dirichlet BC case, which can be interpreted as the generalization 
of the Breitenlohner-Freedman (BF) bound applied to a
general compact 5D warped spacetime. In particular, in a slice of $AdS_5$ the
critical mass is $M^2_{c}=-4k^2 -1/R^2$ and the critical interval distance is
given by $1/R_c^2=|\mu^2|-4k^2$, where $k$ is the $AdS_5$ curvature (the 5D
flat case can be obtained in the limit $k\to 0$, whereas the infinite $AdS_5$
result is recovered in the limit $R\to \infty$). 

\end{abstract}

\maketitle

In recent times the possibility of existence of extra spatial
dimensions~\cite{Rubakov:1983bb,Akama:1982jy,Antoniadis:1990ew,Lykken:1996fj,
Arkani-Hamed:1998rs,Antoniadis:1998ig,Randall:1999ee,Randall:1999vf} 
has become a widely accepted possibility. In a relatively
simple framework, this opens new approaches to deal with some of the puzzles
which still manage to escape our understanding of elementary Particle Physics
and Cosmology.
The space-time geometry can be ``warped'' along the extra coordinate(s) with the
interesting effect of linking hierarchically separated mass scales in an
astonishingly simple setup, as first noted by Randall and Sundrum (RS)~\cite{Randall:1999ee}.
A plethora of phenomenological implications have since been
studied in this context always assuming that the true spacetime background
metric is very close to the simple $AdS_5$ as introduced by RS. When the
static spacetime background is different, one should treat each background case
separately.
Nevertheless one may still try to make general statements and extrapolations 
without specifying exactly the new warp factor considered~\cite{Hirn:2006nt,Delgado:2007ne,Agashe:2007mc}.
This might also be a good laboratory to study the
possible extrapolation or not of such concepts as the holographic
interpretation of 5D warped scenarios in backgrounds other than $AdS_5$.

In this short letter we focus our attention on a real scalar field theory
defined on any warped 5D compact spacetime and such that its 5D mass term is
allowed to be negative\footnote{The same analysis can be carried out
  for vector fields with a negative bulk mass term in the lines of \cite{Batell:2005wa}}.
Naively one can think that instabilities will always occur, but it turns
out that the question depends crucially on the boundary conditions (BC) verified by the
field in the extra compact dimension.
In a pure $AdS_5$ background (when the two brane-boundaries have an infinite separation) 
it is well known that a small enough negative mass term is not
inconsistent with the stability of the system~\cite{Breitenlohner:1982bm,Mezincescu:1984ev}. 
More precisely the mass term $-|\mu^2|$ must verify $-4k^2\leq-|\mu^2|$, where
$k$ is the $AdS_5$ curvature. This bound is generally referred
to as the Breitenlohner-Freedman (BF) bound and when one positions the two 
boundaries at a finite distance one has to treat the bound with care (this was
first addressed in \cite{Delgado:2003tx}).
Also, when the setup involves a generic warped geometry presumably there will also exist a BF type
of bound, allowing well behaved systems around local maxima of the 5D scalar
potential (i.e. with a bulk negative mass squared).

Let's consider a sector of a 5D scenario with a single real scalar field $\phi=\phi(x,y)$
defined by the following
action:
\beq
S_\phi=\int d^4xdy \sqrt{g}\ \left(\frac12 \; \partial^M \phi \;\partial_M \phi
  - {1\over 2 }\mu^2\phi^2\right),
\label{5Dscalaraction}
\eeq
where $y$ represents the extra space coordinate.
The fifth dimension will be treated as an interval and we will concentrate our
attention mainly on separated Neumann and Dirichlet BC's
on the scalar field (which can also be understood as looking for the even and odd
solutions in a $S_1/Z_2$ orbifold), and perhaps comment on separated mixed
BC's if necessary or relevant (when scalar boundary terms are considered).

The background spacetime metric is assumed to take the general form 
\bea
ds^2=e^{-2a(y)}\eta_{\mu\nu} dx^\mu dx^\nu - dy^2\label{warpedmetric}
\eea
where $a(y)$ is a generic warp factor. It is a solution to the static gravitational
background equations resulting from the gravitational sector of
the scenario, which we assume is also stabilized with a fixed separation $\pi R$
between the two boundaries.

Specifically, we are interested in studying the perturbative spectrum of the field
$\phi$ around its trivial vacuum solution $<\phi>=0$, in the special case that 
$\mu^2<0$.

The Euler-Lagrange equation for such perturbations is
\bea
e^{2a}\partial_\mu \partial^\mu \phi - \phi'' + 4 a' \phi' + \mu^2 \phi =0 
\label{5DEulergeneral}
 \eea 
where primes are derivatives with respect to $y$.

Upon separation of variables, equation (\ref{5DEulergeneral}) leads to the
Kaluza-Klein (KK) mode equation
\bea
\phi_y'' - 4a'\phi'_y - (\mu^2-m^2_n e^{2a}) \phi_y=0\label{yeqgeneral}
\eea
where $m^2_n$ is the mass eigenvalue and $\phi_y$ is the profile along the
$y$-direction of the $n$th KK mode. 

\begin{figure}
\center
\includegraphics[width=8cm]{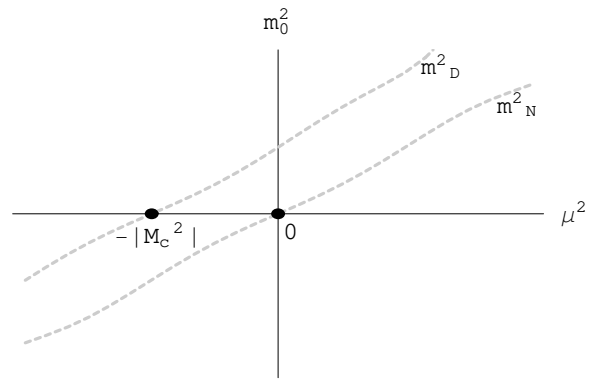}
\caption{Dependence of the lightest eigenvalue for the Dirichlet and Neumann
  problems with respect to the bulk mass $\mu^2$. The big dots correspond to
  the values of $\mu^2$ such that the lightest KK mode is massless. 
  In general, the Neumann zero mode exists only for a vanishing $\mu^2$, while
  the Dirichlet zero mode will exist when $\mu^2$ reaches a critical negative
  value $-|M_c^2|$ which depends on the details of the setup. 
  The curves are model dependent except for the fact that they never cross each other
  and they increase monotonically. The lightest eigenvalue of the mixed BC
  problem will always lie below the Dirichlet curve which therfore sets the best
  bound on $\mu^2$.}
\label{fig1}
\end{figure}

By inspection, it is obvious that $\phi_0(y)=constant\ $ is a massless
solution ($m_0^2=0$) to equation (\ref{yeqgeneral}) when $\mu^2=0$. 
This will actually be the Neumann massless solution for any metric background of the general form given in
equation (\ref{warpedmetric}).
From the general theory of the Sturm-Liouville boundary problem, we know 
that the variation of $m_0^2$ with respect to the bulk mass
parameter $\mu^2$ is always positive, i.e. $\partial m_0^2/\partial \mu^2 >0$
for either Dirichlet or Neumann fields. We also know that there is a strict 
relationship between the eigenvalues of the Neumann and Dirichlet
boundary problems associated with the same Sturm-Liouville operator, namely 
${m^2_0}_N<{m^2_0}_D$. 
This means that when $\mu^2=0$, the lightest
Dirichlet scalar mode will always have a positive mass squared. But it also
proves that to obtain a massless Dirichlet mode, we
require to have a negative bulk mass $\mu^2$, equal to some critical
(negative) mass scale $M^2_c$, i.e. $|\mu^2|=|M_c^2|$ (see Figure \ref{fig1}).
In the case of mixed BC's on the field $\phi$ (i.e. relating the
derivative of the field to its value at the boundaries), there is also a
strict relation between the Dirichlet and the mixed BC eigenvalues
${m^2_0}_{mix}<{m^2_0}_D$.

This means that the bound $M_c^2<\mu^2$ for Dirichlet fields is the best
one can do as far as having a negative bulk mass term.
This simple general result can be seen as a generalization of the
Breitenlohner-Freedman~\cite{Breitenlohner:1982bm,Mezincescu:1984ev} bound for the case of a general warped and
compact 5D setup, applied to fields with either Dirichlet, Neumann or mixed
BC's:

{\it ``Instabilities will always occur in any 5D scalar field theory
defined on a warped interval if it contains a negative bulk mass $\mu^2$ which
is less than the critical mass $M_c^2$ required to obtain massless Dirichlet
excitations. When the bulk mass is above that bound, the type of
boundary conditions will affect the stability or instability of the setup''}.

\begin{figure}[t]
\center
\includegraphics[width=7.5cm]{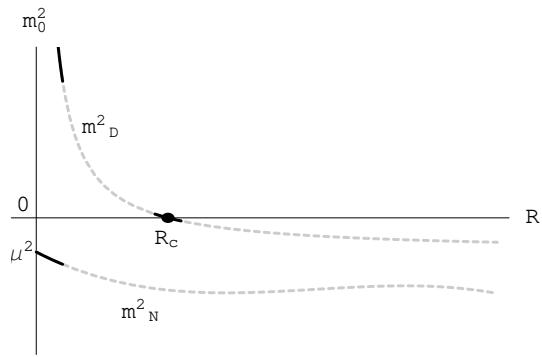}
\caption{Dependence of the lightest mass eigenvalues for the Dirichlet and
  Neumann BC's cases with respect to the size of the interval $\pi R$. 
  $\mu^2$ is negative and fixed, and $R_c$ is the radius of the interval such
  that the Dirichlet lightest mode is massless. The curves are again model
  dependent, except for the fact that ${m_0^2}_D(R)$ is always monotonically decreasing with $R$, it
  vanishes for $R=R_c$ and it has the limit ${m_0^2}_D(R)\to\infty$ as $R\to0$. 
  The Neumann eigenvalue ${m_0^2}_N(R)$ is always below the
  Dirichlet curve and it has a limit ${m_0^2}_N(R)\to \mu^2$ as $R\to 0$.
  For a fixed and negative enough $\mu^2$, there is always a critical
  interval distance $\pi R_c$ such that above it, there will always exist
  instabilities, no matter what boundary conditions one imposes.}
\label{fig2}
\end{figure}

We can also look at things differently, specially perhaps if cosmological
phenomenology is of interest. Instead of studying the dependence on the parameter $\mu^2$, we can
hold it fixed to some negative value and instead treat the distance $\pi R$ between boundaries
as a free parameter so that we can learn about the dependence of the eigenvalues
and eigenfunctions with $R$, while holding everything else fixed.
Again, we invoke another general result from the Sturm-Liouville
theory, only for the Dirichlet case this time,
which states that $\partial {m^2_0}_D/\partial R < 0$ always. This
means that if we fix $\mu^2$ in such a way as to obtain a massless Dirichlet
excitation, for a given boundary distance $\pi R_c$, then we know that for $R>R_c$, we
will have instabilities whereas for $R<R_c$ the system will be well
behaved. Moreover, as $R\to 0$, the lightest Dirichlet eigenvalue will always diverge to
$+\infty$ (see Figure \ref{fig2}). 
In the case of the lightest Neumann eigenvalue, we know that it must
be less than the Dirichlet eigenvalue, and that when $R\to 0$ its value 
approaches the bulk mass $\mu^2$ (see Figure \ref{fig2}). The lightest eigenvalue
corresponding to the mixed BC case is always below the Dirichlet one, and in this
sense again, it is the Dirichlet eigenvalue which sets the tightest bound on
the possible size of the interval in order to avoid instabilities when a
negative bulk mass scalar is considered.

It is very illuminating to consider a simple, yet non trivial example, for which one can define
explicitly both the critical mass and critical radius in a transparent way,
namely a scalar field defined in a slice of $AdS_5$.

\vspace{.2cm}
\begin{center}
{\bf AdS$_5$ CASE\\}
\end{center}

The background spacetime metric 
now contains the warp factor $a(y)=k y$, where the $AdS_5$ curvature $k$
depends on the bulk cosmological constant, adequately tuned with brane tensions
on the two boundaries of the interval\footnote{We assume that some
  mechanism (for e.g.~\cite{Goldberger:1999uk}) fixes and stabilizes the extra dimension,
with negligible backreaction on the metric so that $\sigma(y)=k
y$ remains an acceptable solution.}.
Equation (\ref{yeqgeneral}) becomes here
\bea
 \phi_y'' - 4 k\phi'_y - (\mu^2-m^2_n e^{2ky}) \phi_y=0.\label{yeq}
\eea
Solutions of this equation are well known in terms of Bessel functions 
(see for example \cite{Gherghetta:2000qt}) but it is simpler
to study the conditions to obtain a massless excitation.
When setting $m_0^2=0$ to equation (\ref{yeq}) 
the general solution is then 
\bea
\!\!\!\phi_y(y)\!\!&=&\!\!e^{2ky}\left(A\ \sinh{\sqrt{\alpha}y }+B
  \cosh{\sqrt{\alpha}y}\right), 
\eea
where $\alpha=4k^2+\mu^2$, and $A$ and $B$ are constants. 
When $ \alpha<0$, the hyperbolic functions are simply replaced by
the trigonometric $sin$ and $cos$ functions\footnote{And if $\alpha=0$, the solution is
  $\phi_y(y)=e^{2ky} (Ay+B)$, with $A$ and $B$ constants.}.
From the previous section we already know that for Neumann boundary
conditions, the massless solution corresponds to a constant and exists only
when the bulk mass is zero, i.e. $\mu^2=0$.

The Dirichlet case is more interesting. When $\alpha>0$ it is easy to realize that
there will never exist a massless solution, i.e. the setup is always stable.
On the other hand, when $\alpha<0$ the Dirichlet condition picks up the
trigonometric $sine$ function, i.e. $\phi(y)=A e^{2ky}
\sin{\sqrt{|\alpha|}\ y}\ $, meaning that a massless zero mode solution
can always exist as long as the interbrane distance $\pi R$ is such that $\sqrt{|\alpha|}=1/R$.
The condition on the (negative) bulk mass is therefore $\mu^2=-4k^2-1/R^2=M_c^2$.
From the previous section we know that when $\mu^2<M^2_c$, at least one tachyonic
excitation appears, whereas for $\mu^2>M_c^2$ the Dirichlet system is stable.

The critical distance $\pi R_c$, for a sufficiently negative bulk mass $\mu^2$ is
given by $1/R^2_c=-4k^2 +|\mu^2|$, so that when $R>R_c$, instabilities will
always exist for any type of scalar (Dirichlet, Neumann or mixed). When $-4k^2 +|\mu^2|$
is negative, this indicates that the strict bound on the interval distance does not apply since
Dirichlet excitations will always be stable, but Neumann excitations will be unstable
(for $\mu^2<0$), and mixed BC excitations can be either stable or unstable
depending on their specific BC's~\cite{Delgado:2003tx}.
\vspace{.2cm}
\begin{center}
{\bf \Large Outlook}\\
\end{center}
In the context of a warped and compact extra dimension, we studied the
limits on a negative bulk scalar mass term such that perturbative stability is
maintained. 
With very simple Sturm-Liouville theory techniques, we managed to show that there
will always be a negative mass bound, and pointed out that the best limit
will always correspond to studying the Dirichlet BC case. We also noted that even
in the RS metric, one can obtain a simple and transparent negative mass bound
corresponding to the Breitenlohner-Freedman bound applied to a slice of
$AdS_5$. Namely, any 5D scalar field theory will always be unstable around the
$<\phi>=0$ background if the scalar bulk mass $\mu^2$ violates the bound
$-|\mu^2|>-4k^2-1/R^2$, where $\pi R$ is the interval distance and $k$ is the $AdS_5$ curvature. 
It is however unclear what would be the holographic interpretation of a
bulk scalar field with a negative bulk mass below the original BF bound
$-|\mu^2|>-4k^2$.
In the usual $AdS/CFT$ dictionary a bulk scalar field is interpreted as a 4D
field coupled to an operator with scaling dimension $\Delta= 2+
\sqrt{4+\mu^2/k^2}$.
Now if $-4k^2-1/R^2<-|\mu^2|<-4k^2$, this term becomes imaginary, suggesting that
${\cal{O}}(1)$ boundary corrections might be needed for these fields.

One can also understand the mass bound as a bound on the size of the interval, such
that for a sufficiently negative bulk mass, there will always be a critical size above
which the setup will always be unstable, no matter what BC's are verified by the
fields.
Our original motivation for studying this setup was the search for instabilities of a
Dirichlet scalar field in a warped 5D setup, in an attempt to generalize some
of the results of \cite{Toharia:2007xe,Toharia:2007xf}.
Cosmologically, one could imagine a scenario in which the interval distance $\pi R$ changes on a
cosmological time scale decoupled from the scalar field excitations scale. 
Then, whenever the interval size reaches the critical size,
the scalar sector of the theory will have to undergo a phase
transition, either restoring the scalar potential symmetries or breaking them
(note that no thermal effects are considered here since this is a zero temperature analysis).
It would then become necessary to study the existence and stability of other possible static configurations
of the scalar field sector (coupled now to the gravitational sector), into which the system could decay. In the case
of a flat spacetime (with a decoupled gravitational sector), this was studied
in \cite{Toharia:2007xe,Toharia:2007xf} and the case of
a warped extra dimension is under current investigation~\cite{ttw}.

\vspace{.2cm}
\begin{center}
{\bf \Large Acknowledgments}\\
\end{center}
I would like to thank Kaustubh Agashe, Csaba Csaki, Takemichi Okui, Mark
Trodden, Diana Vaman and James
Wells for discussions, and the KITP center for its hospitality while
finishing this manuscript. This research was supported in part by the National
Science Foundation under Grant No. NSF PHY05-51164.

\vspace{.2cm}
\begin{center}
{\bf \Large Appendix}\\
\end{center}
Let's consider equation (\ref{yeqgeneral}) written in self-adjoint form
\bea
(p\phi')'-q\phi+\lambda w \phi=0 \label{ST}
\eea
where $p(y)=e^{-4a}$, $\ q(y)=\mu^2 e^{-4a}$ and
$w(y)=e^{-2a}$, with $\ \lambda=m_n^2$ being the associated eigenvalue and
$a(y)$ being the generic warp factor of the setup. Note that both $p(y)$ and $w(y)$
are always positive.
This equation is understood as a boundary value problem in the interval $[0,\pi
R]$ and its solutions, when satisfying appropriate boundary conditions, form a complete and orthogonal set. 
The boundary conditions (BC) that we will consider are Dirichlet, Neumann and Mixed
(or Robin) and we will write the associated eigenvalues as $\lambda_D$,
$\lambda_N$ and $\lambda_M$.
A well known result from Sturm-Liouville theory relates the eigenvalues of
different boundary conditions associated to the
same self-adjoint operator, namely
\bea
&&\{\lambda^N_n,\lambda^{M}_n\} < \lambda^D_n
\eea
where $n$ is the index of the solution, which corresponds to the amount of
nodes that the solution has.

It turns out that one can also study the dependence of the eigenvalues $\lambda$
with respect to the parameters of the boundary value problem, such as $R$ or
$\mu^2$ in our case.
One can prove the following very general results verified by the
eigenvalues of different types of BC's: 
\bea {\partial {\lambda}_D\over \partial R}<0, \ \ \ {\partial {\lambda}_N\over \partial \mu^2}>0,\ \
{\rm and}\  \ {\partial {\lambda}_D\over \partial \mu^2}>0,
\eea
where positivity of $p(y)$ is assumed.
For example, let's first write the Raleigh-Ritz
formula for the eigenvalue $\lambda$
\bea
&&\lambda= {1\over N} {\int_0^{\pi R}\Big[p(y)\phi'(y)^2+q(y)\
  \phi(y)^2\Big]\ dy }
\eea
where $N=\int_0^{\pi R} w(y)\phi(y)^2\ dy$.
We can now take the variation of the Neumann eigenvalue $\lambda_N$ with respect to $\mu^2$, holding $R$ fixed and obtain
simply
\bea
{\partial\lambda_N\over \partial \mu^2}={1\over N}\int_0^{\pi R}
 \phi^2 e^{-4a} dy >0.
\eea

This result must be derived with care, since the eigenfunctions $\phi$ depend themselves on the
parameter $\mu^2$. The other positivity results can be proved in a similar
fashion and are known results of the general theory of the Sturm-Liouville problem~\cite{zettl}.


\end{document}